\documentclass[pra,aps,showpacs,superscriptaddress,twocolumn]{revtex4-1}
\usepackage{color} 
\usepackage{times}
\usepackage{amssymb} 
\usepackage{amsmath} 
\usepackage{graphicx}
\usepackage{epstopdf}
\usepackage[english]{babel}
\usepackage{mathrsfs}
\usepackage{textcomp}
\usepackage{amsthm}
\usepackage{amsfonts}
\usepackage{amssymb}
\usepackage{hyperref}
\usepackage{soul}
\usepackage{verbatim}

\newcommand{\ket}[1]{\vert#1\rangle}

\newcommand{\braket}[2]{\langle#1\vert#2\rangle}
\newcommand{\ketbra}[2]{\vert#1\rangle\langle#2\vert}

\newcommand{\modq}[1]{\vert#1\vert^2}
\newcommand{\modulo}[1]{\vert#1\vert}

\begin{document}

\title{Signatures of the single particle mobility edge in the ground state properties of Tonks-Girardeau and non-interacting Fermi gases in a bichromatic potential}

\author{J. Settino}
\affiliation{Dipartimento di Fisica, Universit\`a degli Studi della Calabria,
Arcavacata di Rende, Italy}
\affiliation{INFN, gruppo collegato di Cosenza}
\author{N. Lo Gullo}
\affiliation{Dipartimento di Fisica, Universit\`a degli Studi di Milano,
Milano, Italy}
\affiliation{INFN, sezione di Milano}
\author{A. Sindona}
\affiliation{Dipartimento di Fisica, Universit\`a degli Studi della Calabria,
Arcavacata di Rende, Italy}
\affiliation{INFN, gruppo collegato di Cosenza}

\author{J. Goold}
\affiliation{ICTP, Trieste, Italy}
\author{F. Plastina}
\affiliation{Dipartimento di Fisica, Universit\`a degli Studi della Calabria,
Arcavacata di Rende, Italy}
\affiliation{INFN, gruppo collegato di Cosenza}

\begin{abstract}
We explore the ground state properties of cold atomic gases,
loaded into a bichromatic lattice,  focusing on the cases of
non-interacting fermions and hard-core (Tonks-Girardeau) bosons,
trapped by the combination of two potentials with incommensurate
periods. For such systems, two limiting cases have been thoroughly
established. In the tight-binding limit, the single-particle
states in the lowest occupied band show a localization transition,
as the strength of the second potential is increased above a
certain threshold. In the continuous limit, when the tight-binding
approximation does not hold anymore, a mobility edge is found,
whose position in energy depends upon the strength of the second
potential. Here, we study how the crossover from the discrete to
the continuum behavior occurs, and prove that signatures of the
localization transition and mobility edge clearly appear in the
generic many-body properties of the systems. Specifically, we
evaluate the momentum distribution, which is a routinely measured
quantity in experiments with cold atoms, and demonstrate that,
even in the presence of strong boson-boson interactions, the
single particle mobility edge can be observed in the ground state
properties.

\end{abstract}

\maketitle


\section{Introduction}
\label{sec:intro} Ultra-cold quantum gases, now prepared in a
variety of configurations in laboratories worldwide, have emerged
as ideal candidates for clean and controllable simulation of
condensed matter physics \cite{Bloch:08}. In particular, both
Bosonic and Fermionic atoms can be loaded and manipulated on
optical lattice potentials \cite{Lewenstein:07}. The lack of
thermal phonons, coupled with the tunability of the interactions
by means of Feschbach resonances \cite{Courteille:98}, has allowed
for the detailed study of a multitude of phase diagrams  of
critical many-body systems, both at equilibrium, and away from it
\cite{Polkovnikov2011,Bloch2012}.

Besides the ability to engineer and tune interactions, a
particularly appealing feature of cold atoms, for the simulation
of condensed matter physics, is the possibility to manipulate the
external potential shapes. In this respect, this paper will focus
on many-body systems in a {\it quasi-periodic geometry},
implemented via an external potential that realizes what is known
as Andr\'e-Aubry (AA)
model~\cite{Aubry1980,Harper1955,Jitomirskaya1999,Modugno2009,Albert2010,Aulbach2004,Grempel1982,Ingold2002,Sun2015}.
This is a one dimensional tight binding model on a lattice, with
nearest-neighbor hopping terms and on-site energies given by a
combination of two periodic functions having non-commensurate wave
numbers, which has been shown to display a metal-to-insulator
transition~\cite{Aubry1980,Harper1955}.

The interplay between geometry and interaction in many-body
systems can generate an impressive range of physical phenomena.
For example, the ground state properties of interacting bosons,
subject to a quasi-periodic potential, show a rich phase diagram
at zero temperature~\cite{Roux2008,Roscilde2008,Deng2008},
displaying superfluid, Bose-glass and Mott insulator phases
depending upon the filling of the lattice, the interactions and
the strength of the potential. Interestingly, a mobility edge (ME)
appears when an extension of the AA model is considered, allowing
for longer-range hopping such as next-nearest-neighbor
terms~\cite{Biddle2010,Biddle2011,Wang2013} or a continuous model
(infinite-range hopping)\cite{Boers2007,Diener2001}, or even
interactions~\cite{peliti16}. The extended-to-localized phase
transition of the AA model, within the framework of many body
physics, has been widely investigated both from a theoretical
point of view~\cite{Roth2003,Larcher2009,Gordillo2015,Reichl2016}
and also from an experimental perspective~\cite{Damski2003,Roth2003esp,Fallani2007,Guarrera2007,Fallani2008,Roati2008,Tanzi2013}.

In this work, we ask how the presence of a ME in the single
particle problem affects many-body measurable quantities, and to
what extent the latter can be used to detect it. To this aim, we
study the ground state properties of both non-interacting fermions
and of strongly-interacting bosons in a bichromatic lattice. After
having briefly described the fermion-based representation of the
strongly interacting boson gas in Sec. \ref{sect:tggas}, thus
motivating the choice of looking at both of the species, in
Sec.~\ref{sec:model} we introduce the model considered and recall
the single particle properties, which are crucial to understand
the results in the many-body case. Then, in
Sec.~\ref{sec:fermions} and Sec.~\ref{sec:hcbosons}, we describe
the effect of the delocalization-to-localization transition, and
of the mobility edge on the many-body ground state, for a system
of non-interacting spinless fermions and for a Tonks-Girardeau
gas.

\section{The Tonks-Girardeau gas}
\label{sect:tggas}

Optical lattices allow to create trapping potentials
that are tight enough in the transversal direction to freeze out all
dynamics in these degrees of freedom \cite{Moritz:03}. A gas of $N$
bosons in such a potential can then be approximated by the
one-dimensional Hamiltonian
\begin{eqnarray}
 \label{eq:TG_ham}
 \mathcal{H}_{0}=\sum_{n=1}^N\left[
      -\frac{\hbar^2}{2m}\frac{\partial^2}{\partial x_n^2}
      +V_{ext}(x_{n})\right]
      +g_{1D}\sum_{i<j}\delta(|x_i-x_j|)\;
\end{eqnarray}
where $m$ is the mass of the particles, $V_{ext}$ is the trapping
potential and $g_{1D}$ is a 1D coupling constant which is derived
from the renormalisation of the three-dimensional scattering
process, $g_{1D}=\frac{4\hbar^2 a_{3D}}{ma_\perp}
\left(a_\perp-Ca_{3D} \right)^{-1}$ \cite{Olshanii:98}. Here
$a_\perp$ is the trap width and $C = - \zeta(1/2) \simeq 1.46035$
is a constant. This Hamiltonian describes an inhomogeneous
Lieb-Liniger gas, which in the strongly repulsive limit,
$g_{1D}\rightarrow\infty$, can be solved by using a mapping to an
ideal and spinless fermionic system \cite{Girardeau:60}. The
essential idea of this mapping is that one can then treat the
interaction term in Eq.~(\ref{eq:TG_ham}) by replacing it with a
boundary condition on the allowed Bosonic wave-function
\begin{equation}
  \label{eq:constraint}
  \Psi_B(x_1,x_2,\dots,x_n)=0\quad \mbox{if} \quad |x_i-x_j|=0\;,
\end{equation}
for $i\neq j$ and $1\leq i\leq\ j\leq N$.  This is the hard core
constraint, which says that no probability exists for two
particles ever to be at the same point in space.

Such a constraint is automatically fulfilled by calculating the wave-function using a
Slater determinant
\begin{equation}
  \Psi_F(x_1,x_2,\dots,x_N) =\frac{1}{\sqrt N!}
                             \det_{(n,j)=(0,1)}^{(N-1,N)}\psi_n(x_j)\;,
\label{eq:psiF}
\end{equation}
where the $\psi_n$ are the single particle eigenstates of the ideal
system. This, however, leads to a fermionic rather than bosonic
symmetry, which can be corrected by a multiplication with the
appropriate unit antisymmetric function~\cite{Girardeau:60}
\begin{equation}
  A=\prod_{1\leq i < j\leq N} \mbox{sgn}(x_i-x_j)\;,\\
  \label{eq:mapFB}
\end{equation}
to give $\Psi_B=A \Psi_F$. Once the single particle eigenstates of
the system in the external potential of interest are known then
the many-body properties of both the free Fermionic and hard-core
Bosonic systems can be investigated.


\section{Single particle problem}
\label{sec:model} Let us consider the time independent
Schr\"{o}dinger equation for a particle in an external potential:
\begin{equation}
\label{eq:sch}
\left[-\frac{\hbar^2}{2m}\frac{\partial^2}{\partial x^2}
      +V_{ext}(x)\right]\psi_{n}(x)=e_n\psi_{n}(x).
      \end{equation}
In what follows, we will consider an external potential which
describes a bichromatic lattice,
\begin{equation}
\label{eq:pot} V_{ext}=V_1 \sin^2 (k_1 x) + V_2 \sin^2 (k_2 x) \,
.
\end{equation}
Although any irrational number would work as well, to be specific
we will take $\frac{k_1}{k_2}=\frac{1+\sqrt{5}}{2}= \tau$, the
golden ratio, and assume $V_1>V_2$. For $V_2=0$ and $V_1\ge 5
E_{R_{1}}$ it is possible to resort to the so called tight-binding
(TB) limit to approximately describe the system. Here
$E_{R_1}=\hbar^2 k_1^2/(2 m)$ is the recoil energy, giving an
estimation of the energy at which the potential with modulation
$V_1$ opens the first gap of width $\propto V_1$ in the otherwise
gapless free particle spectrum (for $V_1=V_2=0$). The above
condition therefore ensures that all particles with energy
$E<E_{R_1}$ do not have enough energy to overcome the first gap
and, therefore, that they are confined in the lowest band of the
potential. In this limit, the properties of the system are
dominated by the external potential and it is possible to rewrite
the single particle Hamiltonian in Eq.~(\ref{eq:sch}) in terms of
states $\ket{i}$, localized around the minima of the potential,
whose wave functions $w_i(x)=\braket{x}{i}$ are the so called
Wannier functions. In the presence of the second potential
$V_2<V_1$, and in the TB limit, the continuum model described by
Eq.~(\ref{eq:sch}) can be mapped into the so called Aubry-Andre
(AA) model:
\begin{equation}
\label{eq:AAh}
\hat H =\Delta \sum_j \cos(2 \pi \tau j) \ketbra{j}{j}- J \sum_j
( \ketbra{j+1}{j}+ \ketbra{j}{j+1}).
\end{equation}
The first term on the r.h.s., proportional to $\Delta$, accounts
for the on-site energy, whereas the second one, proportional to
the hopping constant $J$, is responsible for nearest-neighbor
tunnelling between adjacent sites. Both $\Delta$ and $J$ depend
upon the choice of the set of Wannier functions, which in turn
depend upon the first potential only if the condition $V_2\ll V_1$
is satisfied.

The AA model has been widely studied from different points of
view. For what we are concerned here, it is worth recalling that
the AA model shows a delocalized-to-localized (or
metal-to-insulator) transition at $\Delta/J = 2$. This point marks
the change from a delocalized phase ($\Delta/J<2$), in which all
of the eigenstates have an extended character with a corresponding
absolutely continuous spectrum, to a phase where all states are
localized and the spectrum is discrete~\cite{Jitomirskaya1999}. As
for the many-body properties of this system, it has been predicted
numerically and verified experimentally that bosons with the
addition of on-site interaction in the AA model enjoy a
particularly rich phase diagram, which includes a superfluid to
Bose-glass transition at low filling, and also a Mott insulator
phase for higher filling and interaction strength.

In the continuum, outside the range of validity of the TB
approximation, it is known that the sharp delocalized-to-localized
transition of the lowest energy band transforms into a mobility
edge, whose position in energy changes with
$V_2$~\cite{Biddle2010,Biddle2011}. Our aim is to study in detail
how this crossover from the discrete to the continuum behavior
occurs, and to show that signatures of this transition are
displayed in the many-body properties of both non-interacting
fermions and strongly-interacting bosons. We will therefore always
work with the continuous model of Eq.~\eqref{eq:sch} and move from
the discrete to the continuous limits by changing the strength of
the main potential $V_1$. For each set of parameters $\{V_1,V_2\}$
we have numerically solved the eigenvalue problem given by
Eq.~\eqref{eq:sch} via a fifth order Matrix Numerov Method
\cite{Pillai2012}, considering systems with $N=100$ lattice sites
and total length $k_1 L=100\pi$.


\subsection{Mobility edge in the single particle problem}
\label{sec:spp} 
\begin{figure}[h]
\includegraphics[width=7cm]{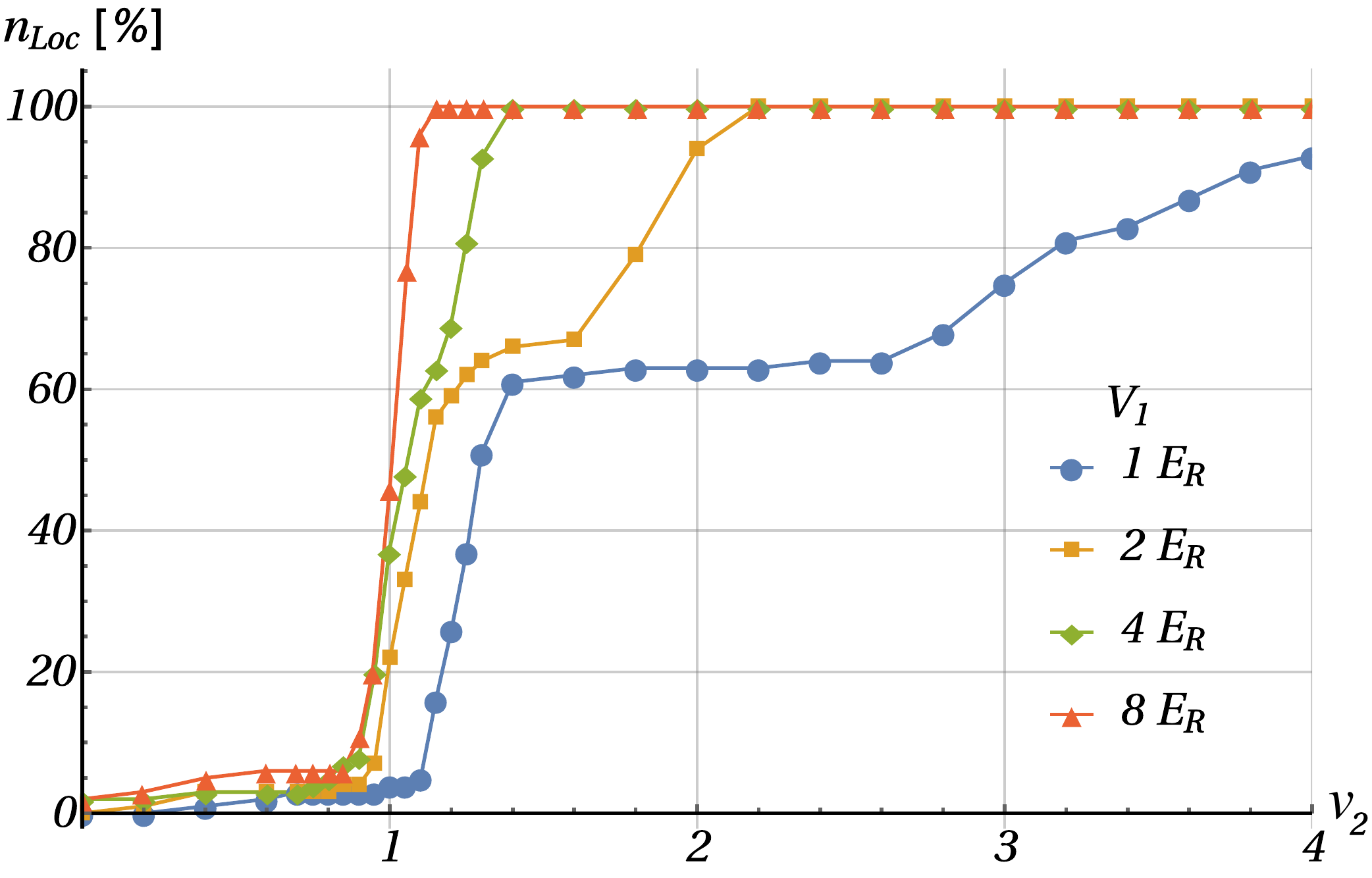}
\caption{(Color online) Number of localized states as a function
of $v_2$ for different values of $V_1$. An eigenfunction is
considered to be localized if its IPR is greater than $1/5l$ where
$l$ is the distance between two neighboring sites. For the system
considered here, $5 l$ corresponds to the $5\% $of the whole
lattice length $L$.} \label{fig:nLoc}
\end{figure}

\begin{figure}[h]
\includegraphics[width=8cm]{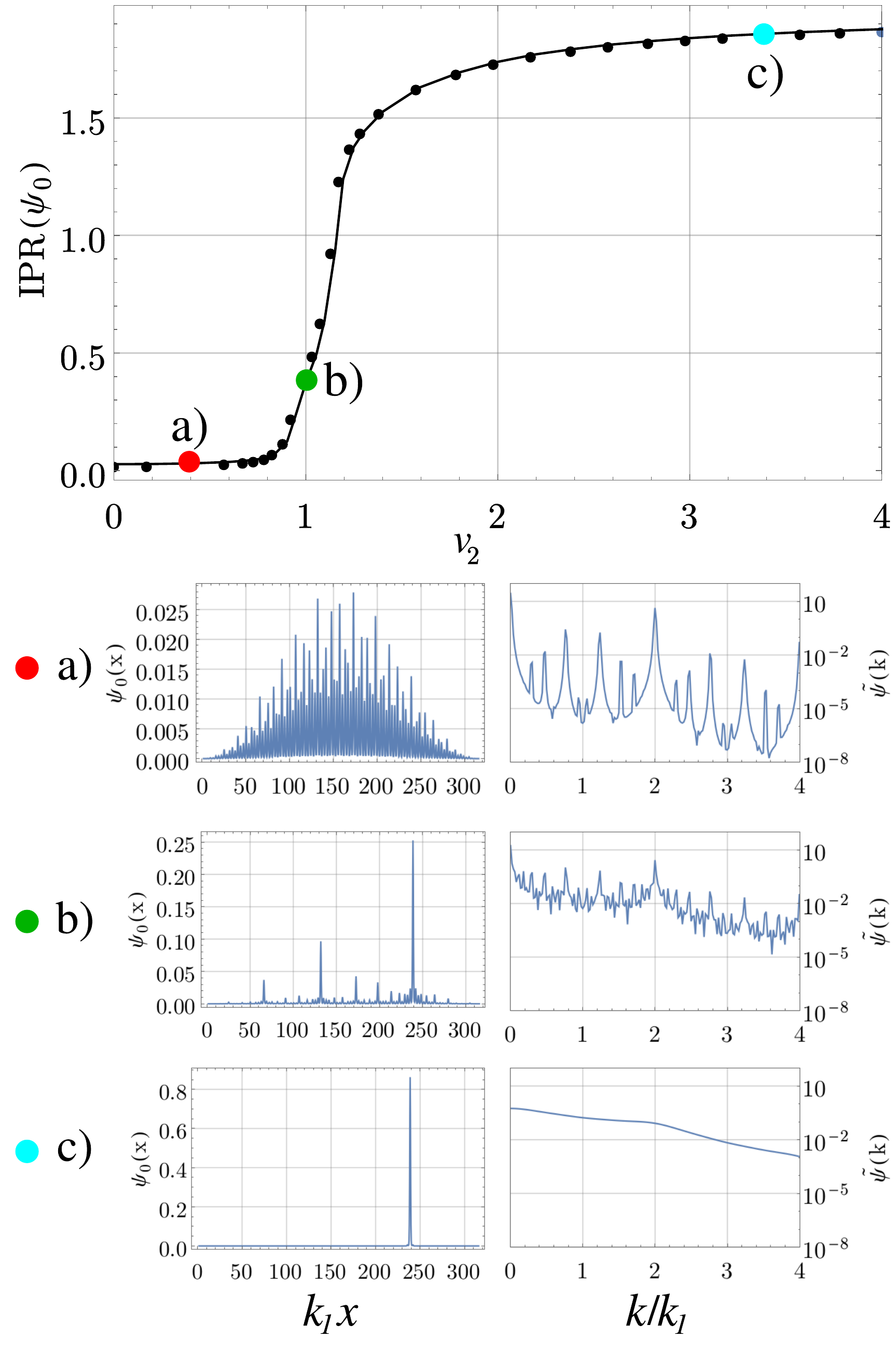}
\caption{(Color online). (Top) Normalized IPR of the single
particle ground state of the system as a function of $v_2$ for
$V_1=8E_R$. (Bottom) Single particle ground state and its Fourier
Transform in the delocalized, critical and localized regions
corresponding to the values of $v_2$ reported in red, green and
cyan colors, respectively, in the top figure.} \label{fig:ipr3}
\end{figure}

In the discrete model, all of the eigenfunctions
of the Hamiltonian of Eq.~\eqref{eq:sch} in the TB regime are
either extended or localized, depending on wether the value of
$V_2$ is below or above a certain characteristic value $V_2^t$. On
the other hand, if $V_1 <5 E_R$ a Mobility Edge (ME) appears such
that, for a fixed value of $V_2$, states with energy lower then
the ME are localized whilst the others are delocalized
\cite{Boers2007}. The ME is found at higher energies for
increasing $V_2$. It is possible to obtain an estimation of the
localization threshold $V_2^t$ by calculating explicitly $\Delta$
and $J$ of the AA model -- as a function of $V_1$ and $V_2$ -- and
inverting the condition $\Delta/J=2$, by solving for $V_2$. Using
the numerical estimations given in \cite{Modugno2009}, we have
obtained
\begin{equation}\label{eq:V2T}
 V_2^t=2 E_{R_2} {14.9752 e^{\frac{0.381966}{\sqrt{V_1/E_{R_1}}}-2.07 \sqrt{V_1/E_{R_1}}}} ({V_1/E_{R_1}})^{0.98}
\end{equation}
In the following we will use the rescaled quantity
$v_2=V_2/V_2^{t}$ in order to compare systems with different $V_1$
and $V_2$. This guarantees that the transition point in the TB
limit always occurs at $v_2=1$. However, it is important to notice
that the computation of $V_2^t$ has been done, and it is
meaningful only in the TB limit. As we will also consider
parameters for which the TB approximation does not hold, then the
value $v_2=1$ will lose its importance and its role of transition
point. To quantitatively discuss the transition in the general
case, we will employ the Inverse Participation Ratio
$\text{IPR}(\psi) = \int \modulo{\psi(x)}^4 dx / \int
\modq{\psi(x)} dx$ for the eigenfunctions of the first energy band
of the Hamiltonian. This quantity measures the inverse of the
average spatial region occupied by the eigenfunction. We will
consider an eigenfunction to be localized if its IPR is larger
than $1/(5 l)$ where $l$ is the distance between two neighbor
lattice sites. Fig.~\ref{fig:nLoc} shows the number of localized
states as a function of $V_2$ for different values of $V_1$: in
the TB regime (upper, red curve) the transition is sharp, whereas
in the continuum there is an ME, as witnessed by the plateaux in
Fig.~\ref{fig:nLoc}, that correspond to gaps inside the first
energy band. Moreover, as anticipated above, Fig.~\ref{fig:nLoc}
shows that the estimation of the transition point $V_2^{t}$ given
in Eq.~\eqref{eq:V2T} fails when the TB description does not
provide a good approximation (see, e.g. the lower curve,
corresponding to $V_1=E_R$). We finally show in
Fig.~\ref{fig:ipr3} the $\text{IPR}$ of the ground state wave
function, normalized to one lattice length, together with the
function itself and its its Fourier transform for various values
of $v_2$.

In the remainder of the paper we will show how both the sharp
transition, occurring in the TB limit, and the appearance of the
ME affect the many-body properties of non-interacting spinless
fermions and Tonks-Girardeau bosons. It is known that the
excitation spectrum of the Tonks-Girardeau model is the same as
that of non-interacting spinless fermions and that all local
quantities are the same for the two systems. On the other hand,
non local quantities, such as the momentum distribution, are
different and reveal the fermionic and bosonic nature of the two
systems. More importantly, we shall see that the different
statistical nature of the two kind of particles also shows up in
the way the localization transition and the appearance of the ME
manifest themselves in the momentum distributions.


\section{Fermions in a bichromatic lattice}
\label{sec:fermions} In this section, we consider the effect of
the ME of the single-particle spectrum on the ground state
properties of a system of $N$ non-interacting fermions loaded into
the bichromatic lattice, and, more generally, discuss the
signatures of the transition from discrete to continuum, occurring
as $V_1$ is decreased.

The many-body wavefunction describing a system of $N$
non-interacting spinless fermions is given by a Slater determinant
of single particle states (see Eq.~\eqref{eq:psiF}). Below, we
focus on the reduced single particle density matrix (RSPDM)
\begin{equation}
\rho_F(x,y)=\int dx_2 ... dx_N \Psi_F^*(x,...,x_N)
\Psi_F(y,...,x_N) \, ,
\end{equation} whose knowledge is sufficient to evaluate the expectation values of
all single-particle operators. Its Fourier transform gives the
momentum distribution (MD),
\begin{equation}
n_F(k)=\frac{1}{2\pi} \int dx dy \exp^{i k (x-y)} \rho_F(x,y),
\end{equation}
which is directly measurable in cold atom experiments. The
eigenfunctions ($\phi_i(x)$) and eigenvalues ($\lambda_i$) of the
RSPDM, defined by $\int dx \rho_F(x,y)\phi_i(x)=\lambda_i
\phi_i(y) $, and normalized such that $\sum_i \lambda_i = 1$, are
the so called {\it natural orbitals} and their populations,
respectively. For a non-interacting Fermi gas, as one could have
expected, there are only $N$ non-vanishing eigenvalues, which are
all equal to one, and the corresponding eigenvectors coincide with
the first $N$ lowest single particle energy states. A similar
analysis for a Tonk-Girardeau gas will prove to be much less
trivial.

The RSPDM and MD of a non-interacting fermion gas in its ground
many-body state can be expressed via the single particle energy
eigenstates: $\rho_F(x,y)=\sum_{j=1}^N \psi_j^*(x)\psi_j(y)$ and
$n(k)=\sum_{j=1}^N \modq{\tilde \psi_j(k)}$ where $\tilde
\psi_j(k)$ is the Fourier transform of $\psi_j(x)$
\cite{Pezer2007}.

\begin{figure}[h]
\includegraphics[width=7cm]{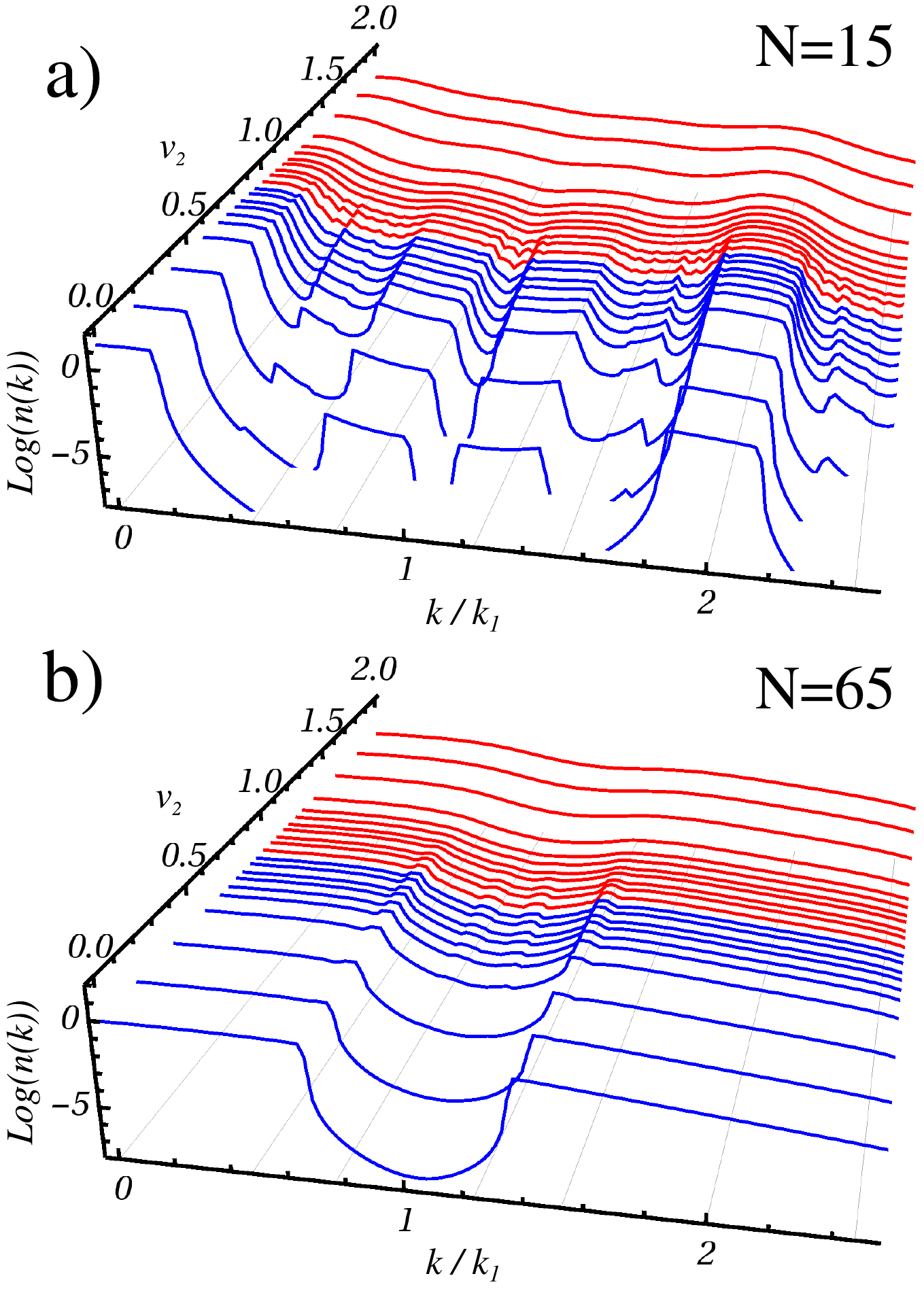}
\caption{(Color online). Momentum distribution $n(k)$ as a
function of $v_2$ for a system of non-interacting fermions with
$V_1=8E_r$ (TB limit). The two panels refer to a different number
of fermions: a) N=15 and b) N=65. Blue and red curves are for
$v_2<1$ and $v_2>1$ respectively.} \label{fig:momDistTB}
\end{figure}

We analyze the TB limit first, where, interestingly enough, the MD
offers a signature of the localization transition inherited from
the single particle properties. On the delocalized side, the
Fourier transform of each single particle state displays peaks at
the wave numbers $k(m,n)= 2 m k_1 \pm 2 n k_2 $ with $m, n \in
\mathbb{Z}$, and the MD shows several Fermi-Dirac-like flat
structures due to the occupation of states with nearby momentum
peaks. This is shown in Fig.~\ref{fig:momDistTB}, where explicit
reference to gases of $N=15$ and $N=65$ fermions is  made. In the
delocalized region ($v_2<1$), indeed, some nearly flat regions
appear in the MD. They are centered at different $k(n,m)$'s, and
are due to the fact that each single particle wavefuction
contributes with at least two momenta (but in general more, for
higher energy states); these momenta pile up in the total $n(k)$
to give a sequence of nearby peaks forming these Fermi-Dirac-like
(almost) flat regions, whose width is proportional to the number
of particles.

On the other hand, again in the TB regime, but now in the
localized domain, the MD suddenly smoothens for $v_2>1$, due to the
fact that single particle states are all localized, so that their
Fourier transform flattens over in momentum space. Before shifting
our attention to the continuum case, it is instructive to discuss
the nature and origin of the structures that appear at the edges
of the flat regions. They are particularly well visible in the
case of $N=65$. Conversely, for $N=15$, they only show up as small
peaks, whose height increases with increasing $v_2$ until the
transition point $v_2=1$ is reached, where they disappear leaving
structure-less bumps.

To understand why this happens, we recall that the addition of the
second potential leads to a fragmentation of the energy spectrum
into sub-bands separated by gaps whose width depends, at first
order in a perturbation analysis, on the amplitudes of the Fourier
transform of the potential itself. Furthermore, the sub-bands tend
to flatten out as the potential is increased. This implies that
the density of states increases at the sub-band edges and more
particles can be accommodated there. As a result, the structures
at the edges of the flat regions become better and better defined
as the number of particles increases.

\begin{figure}[h]
\includegraphics[width=7cm]{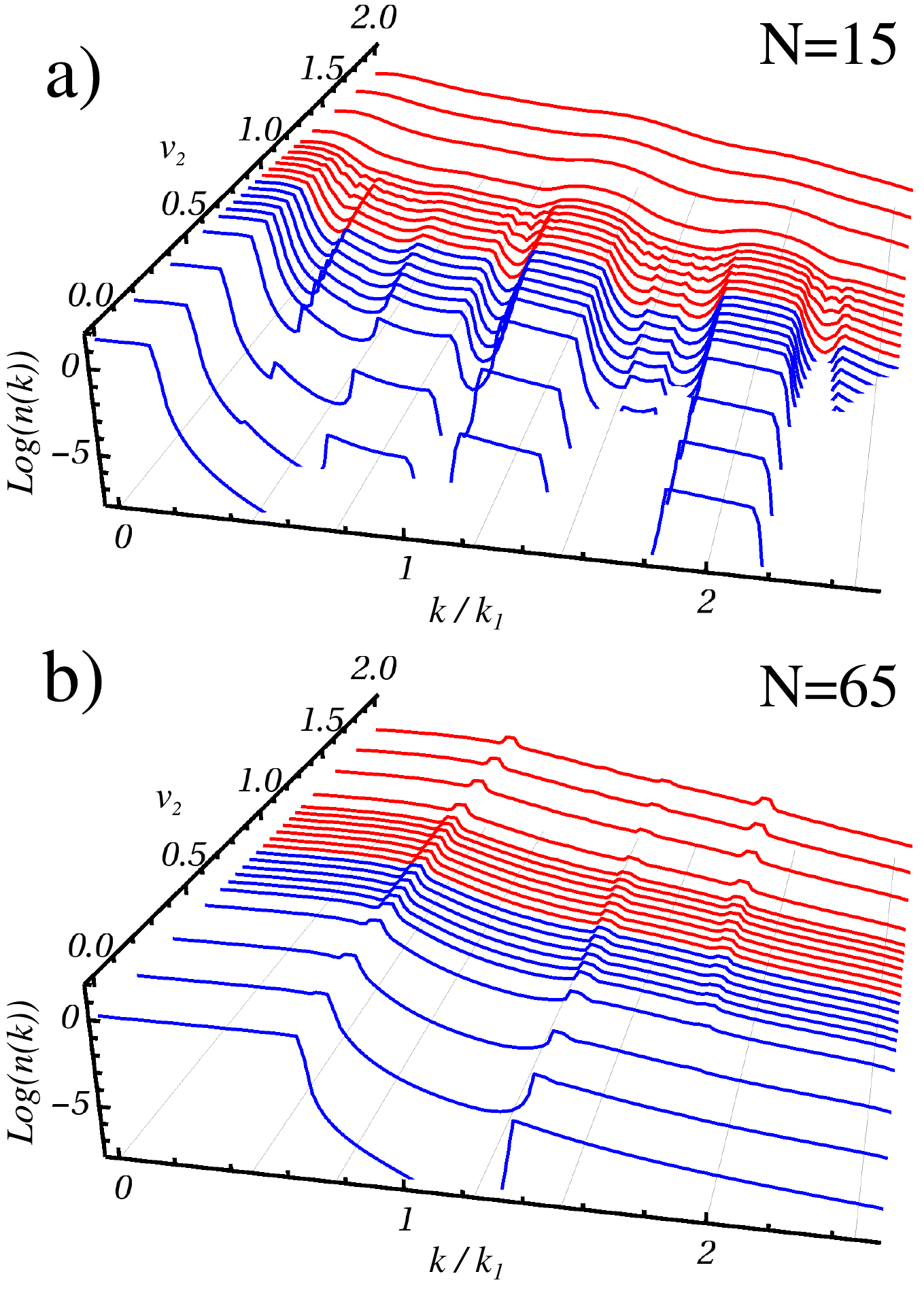}
\caption{(Color online). $n(k)$ as a function of $v_2$ for a
system of non-interacting fermions with $V_1=8E_r$. The same
parameters and coloring as Fig. \ref{fig:momDistTB} is used. A
persistence of the edge-structures well beyond $v_2=1$ is observed
for $N=65$, but not for $N=15$ as, for these parameters, the first
$15$ single particle states remain below the ME.}
\label{fig:momDistCNT}
\end{figure}

In the continuum, namely $V_1\approx E_R$, the
delocalized-to-localized transition turns into the appearance of
an ME. Indeed, Fig. \ref{fig:momDistCNT} b) shows that it is
possible to observe structures in the MD of an $N=65$ fermion gas
at any value of $v_2$, and in particular well beyond the
transition point $v_2=1$, which used to mark a sharp transition in
the TB approximation. The persistence of such structures is a
signature of the existence of occupied single particle states,
which remain delocalized beyond $v_2=1$.

However, this seems not to be the case if one looks at
Fig.~\ref{fig:momDistCNT} a), where $n(k)$ is shown for $N=15$
fermions, instead. Here, the structures of the MD quickly
disappear when increasing $v_2$ beyond $1$. In fact, as we
considered a smaller number of fermions in this case, less states
are occupied, and all of them become localized for $v_2 >1$
(giving rise to a quick smoothing of $n(k)$). The comparison
between Fig.~\ref{fig:momDistCNT} a) and Fig.~\ref{fig:momDistCNT}
b), therefore, shows that, in the same band, there exist both
localized single particle states (at low energy) and delocalized
ones (at higher energy), which is a clear manifestation of the ME.

In order to characterize quantitatively this phenomenon, we resort
to an approach already used in Ref.\cite{Macia1999} to study the
degree of delocalization of phonon modes in quasi-crystalline
structures. The idea behind it is to evaluate the area under the
MD peaks, in order to quantify the total power spectrum coming
from coherent sources, which in our case correspond to delocalized
single particle states. Therefore, we have evaluated the total
area $I_d$ below the edge-peaks in $n(k)$ after removing the
background. On the delocalized side ($v_2<1$) $I_d$ increases as a
function of $v_2$ due to the appearance of new peaks, induced in
the single particle states by the second potential, when the
transition point is approached. Moreover, for small numbers of
fermions ($N = 15$) such an increase is approximately linear, up
until saturation is reached. This behavior stems from the fact
that only the lower energy eigenstates are occupied and,
therefore, large values of the second potential are needed to make
them develop a spatial structure, which involves more momenta. On
the other hand, for $N = 65$, saturation occurs well before, due
to the fact that higher energy eigenstates are occupied even at
small values of $v_2$,  which contribute to $n(k)$ with more
momenta and, therefore, more peaks.

\begin{figure}[h]
\includegraphics[width=7cm]{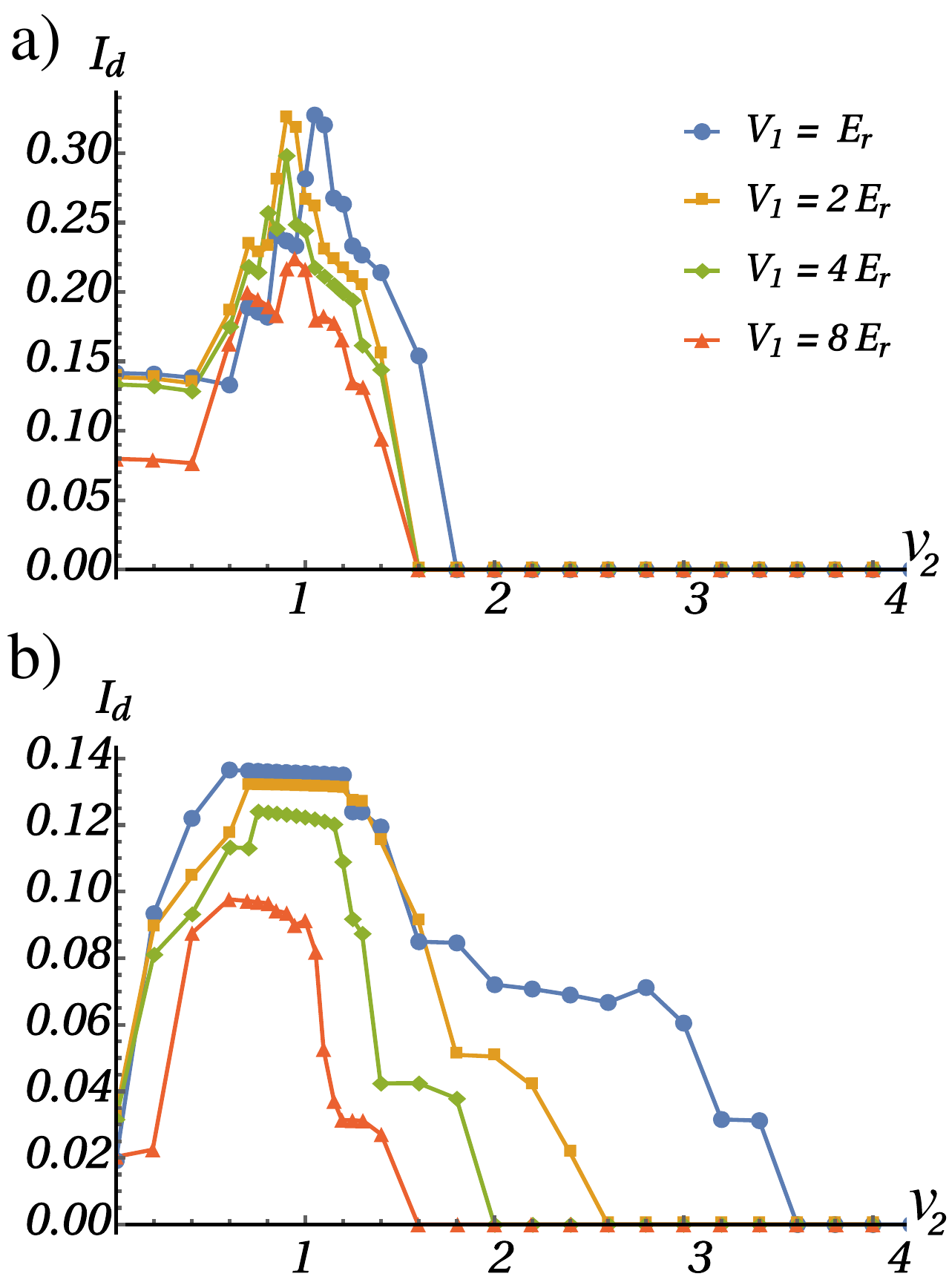}
\caption{(Color online). $I_d$ for a) $N_p=15$ and b) $N_p=65$ fermions.}
\label{fig:DescretePeaksMomDistF}
\end{figure}

When the localized eigenstates start to be occupied, $I_d$
decreases due to the smoothing of the MD profile. In the
TB-regime, $I_d$ suddenly goes to zero for $v_2 > 1$, regardless
of the number of fermions, as all of the eigenstates suddenly
localize. On the other hand, in the continuous limit with $V_1
\approx E_r$, we observe the appearance of an ME as $I_d$
decreases for $v_2 > 1$ and reaches a plateaux (blue and yellow
curves in Fig. \ref{fig:DescretePeaksMomDistF} b). The plateaux,
in particular, witness the ME moving through the band towards
higher energies.

\section{Tonks-Girardeau gas in a bichromatic lattice}
\label{sec:hcbosons} In Sec.\ref{sec:fermions} we have seen that
signatures of the ME appear in the momentum distribution of the
the RSPDM of a many-body system of spinless fermions. In this
section we look at a strongly-interacting boson gas (i.e., the
Tonks-Girardeau gas), whose ground state properties can be related
to the fermionic ones via the prescription outlined in
Sec.\ref{sect:tggas}.

The many body spectrum of an hard-core boson system is, in fact,
the same as that of the noninteracting fermion gas, loaded into
the same optical potential $V (x)$. Moreover, it is possible to
show that the mapping from hard-core bosons to non-interacting
fermions leaves all {\it local} quantities unchanged; for
instance, given the many-body wavefunction for $N$ hard-core
bosons, the density of bosons is the same as that of $N$ non
interacting fermions. Conversely, non-local properties, such as
correlation functions, are different in the two cases. For this
reason we expect the MD, which is derived from the off-diagonal
entries of the RSPDM, to be markedly different from that observed
in Sec.\ref{sec:fermions}, because of the presence of spatial
coherences, typical of a boson gas.

In Ref.~\cite{Pezer2007}, it has been shown that the RSPDM of $N$
hard-core bosons can be obtained by the single particle
eigenstates of the equivalent non-interacting fermion problem as:
\begin{equation}
\rho_B(x,y)=\sum_{i,j=1}^N \psi_i^*(x) A_{i j}(x,y) \psi_j(y)
\label{eq:RSPDMbosons}
\end{equation}
where $A_{i j}(x,y)=(\mathbf{P}(x,y)^{-1})_{j i} \det
(\mathbf{P}(x,y))$ and the $N \times N$ matrix $\mathbf{P}$ is
given by $P_{i j} (x,y)= \delta_{i j} - 2\int_x^y dx' \psi_i^*(x')
\psi_j(x')$. As in the fermionic case, the many-body MD is
obtained by a Fourier transform of the RSPDM,
$n_B(k)=\frac{1}{2\pi} \int dx dy \exp^{i k (x-y)} \rho_B(x,y)$.

The MD is conveniently expressed in terms of the natural orbitals
and their corresponding eigenvalues, $n(k)=\sum_{j=1}^N \lambda_j\modq{\tilde
\phi_j(k)}$. Here, $\tilde \phi_j(k)$ is the Fourier transform of
the $j$-th natural orbital, given by the eigenfunction of the
RSPDM: $\int dx \rho_B(x,y)\phi_j(x)=\lambda_j \phi_j(y) $.

As expected, the bosonic MD is markedly different from that of the
non-interacting fermion case. Due to the their bosonic nature, at
zero temperature the particles would tend to occupy the single
particle modes with lowest energy. On the other hand, due to the
very strong repulsion between two bosons when their spatial
overlap is large, particles tend to occupy states in order to
lower their overlap. The interplay between these two effects is
the key mechanism that explains the behavior of the MD for
strongly interacting bosons, which we will now analyze.
\begin{figure}[h]
\includegraphics[width=7cm]{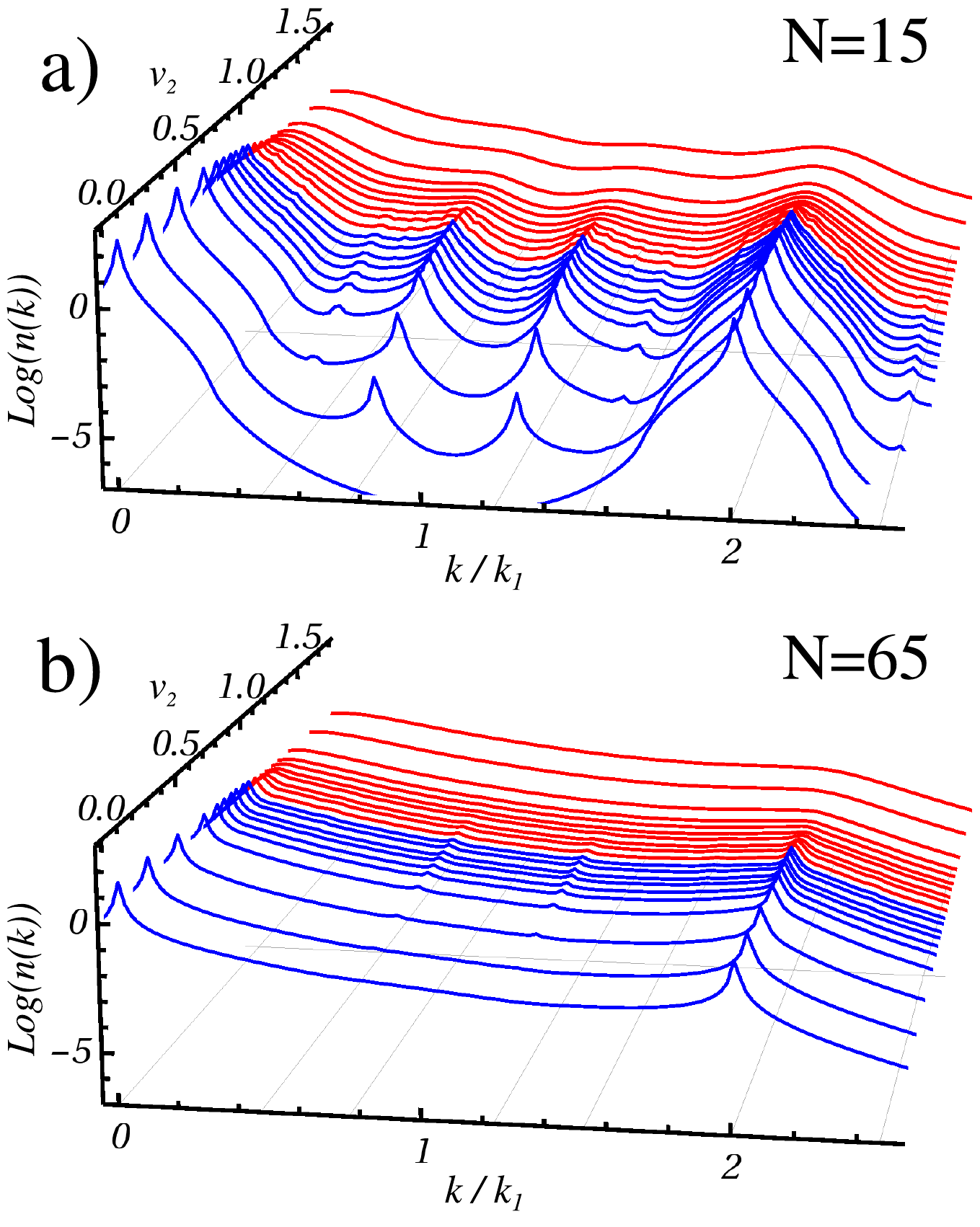}
\caption{(Color online). Momentum distribution $n(k)$ as a
function of $v_2$ for a system of strongly-interacting bosons with
$V_1=8E_r$ (TB limit). Different figures refer to a different
number of bosons: a) N=15 and b) N=65. Blue and red curves are for
$v_2<1$ and $v_2>1$ respectively.} \label{fig:momDistHCBTB}
\end{figure}

\begin{figure}[h]
\includegraphics[width=7cm]{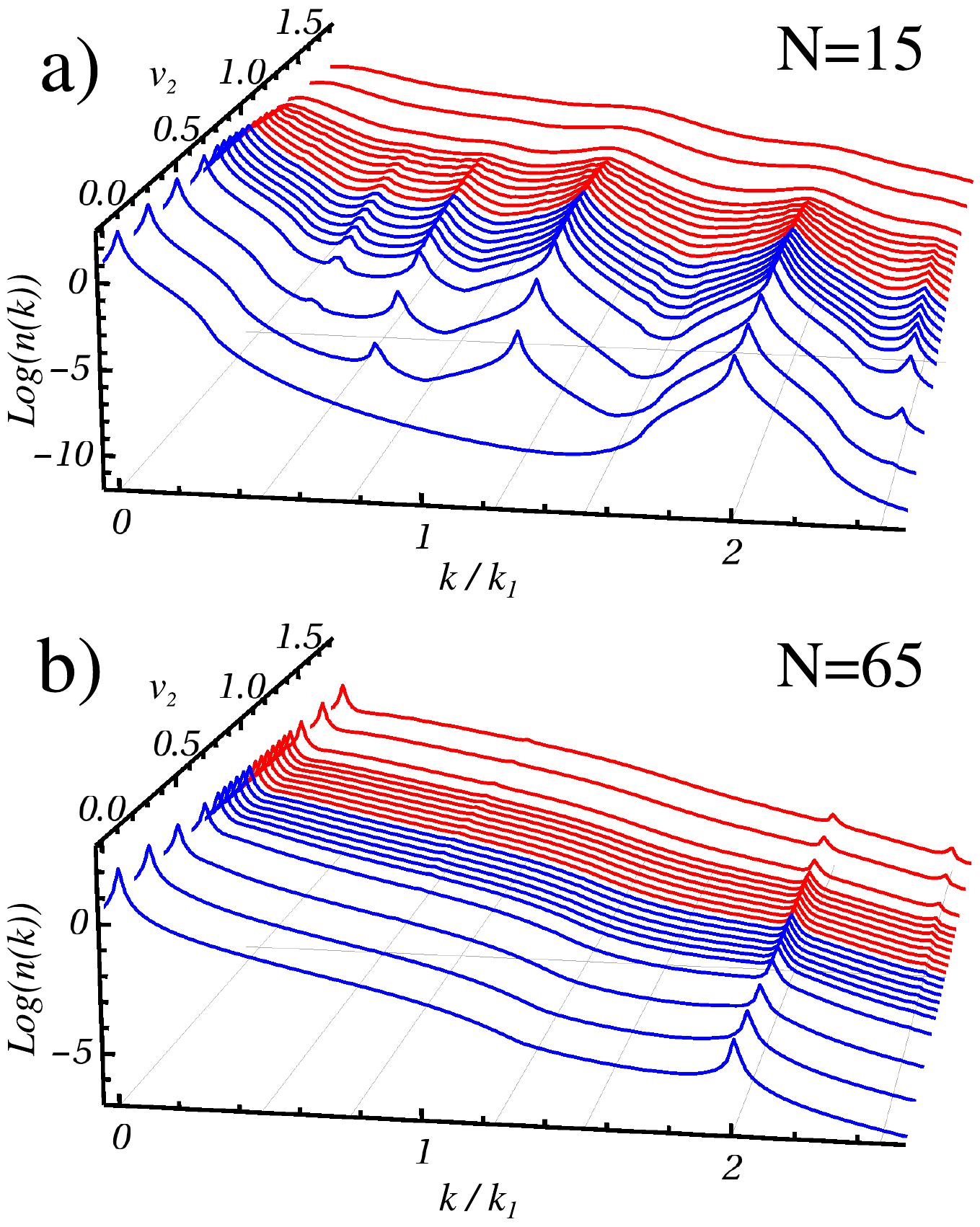}
\caption{(Color online). Momentum distribution $n(k)$ as a
function of $v_2$ for a system of strongly-interacting bosons with
$V_1=E_r$. Different figures refer to a different number of
bosons: a) N=15 and b) N=65. Blue and red curves are for $v_2<1$
and $v_2>1$ respectively.} \label{fig:momDistHCBCNT}
\end{figure}

On the delocalized side $v2 < 1$, the sharp peaks of the
noninteracting fermion case are replaced by broad peaks centered
at specific momenta of the form $k(m,n)=2m k_1+ 2 n k_2$, induced
by the trapping potential.

Indeed, as stated above, bosons would want to occupy a low energy
mode, but not all can occupy the same one, as this would lead to a
large overlap between their wavefunctions and, therefore, to an
increase of the total energy. This effect is taken into account by
the presence of the factors $A_{ij}$ in the RSPD matrix, as built
from single particle eigenfunctions in Eq.\ref{eq:RSPDMbosons}.
Such a behavior gets more and more pronounced with increasing the
number of bosons, as witnessed by panel a) and b) in Fig.
\ref{fig:momDistHCBTB}, where we display the MD for $V_1=8E_r$ in
a system of $N=15$ and $N=65$ bosons as a function of $v_2$. In
the first case [Fig. \ref{fig:momDistHCBTB} a), $N=15$], two new
peaks appear as the second potential is switched on, which are
very well pronounced and persist up until the transition point
$v_2=1$ is reached. For $v_2>1$, instead, the peaks broaden,
$n(k)$ smoothen out and any structure is lost.

In the second case [Fig. \ref{fig:momDistHCBTB} b), $N=65$], the
two peaks emerge from an already large background, which is due to
the large number of bosons in the system that tend to occupy more
states. This implies that the tails of the main peaks, due to the
main potential $V_1$, are quite high. Notwithstanding the fact
that they are immersed in these tails, they are still visible,
therefore witnessing the spatial coherence of delocalized modes.

\begin{figure}[h]
\includegraphics[width=7cm]{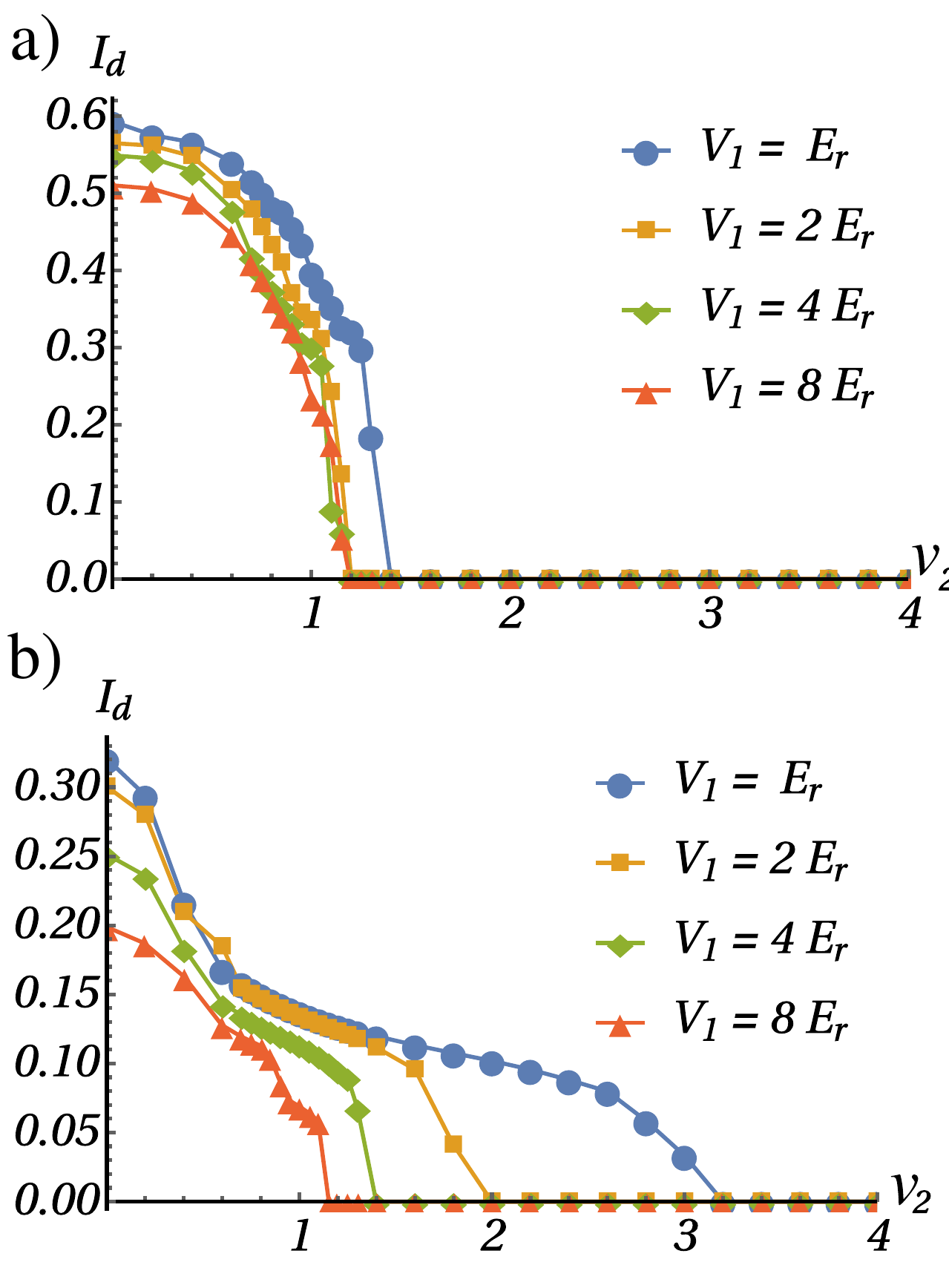}
\caption{(Color online). Area under the discrete peaks, $I_d$, as
a function of $v_2$ for various values of $V_1$, and for a)
$N_p=15$ and b) $N_p=15$ bosons.}
\label{fig:DescretePeaksMomDistB}
\end{figure}

To better highlight such structures, we performed a peak-area
analysis, similar to that discussed for fermions. Fig.
\ref{fig:DescretePeaksMomDistB}b shows that for $V_1<4E_r$ the
areas under the descrete part of the MD does not go to zero at the
transition point, confirming the presence of the ME.

Unlike the non-interacting fermion case, where the occupancy of
each natural orbital was either $0$ or $1$, here it is meaningful
to look at their distribution and at the entropy of the RSPDM
given by $S(\rho_B)=-\sum_i\lambda_i \ln \lambda_i$. 
For $N$ strongly interacting bosons at the absolute zero, 
we should not always expect that the first $N$ energy levels are occupied,
and, correspondigly, $S(\rho_B)\neq\ln N$.
\begin{figure}[h]
\includegraphics[width=7cm]{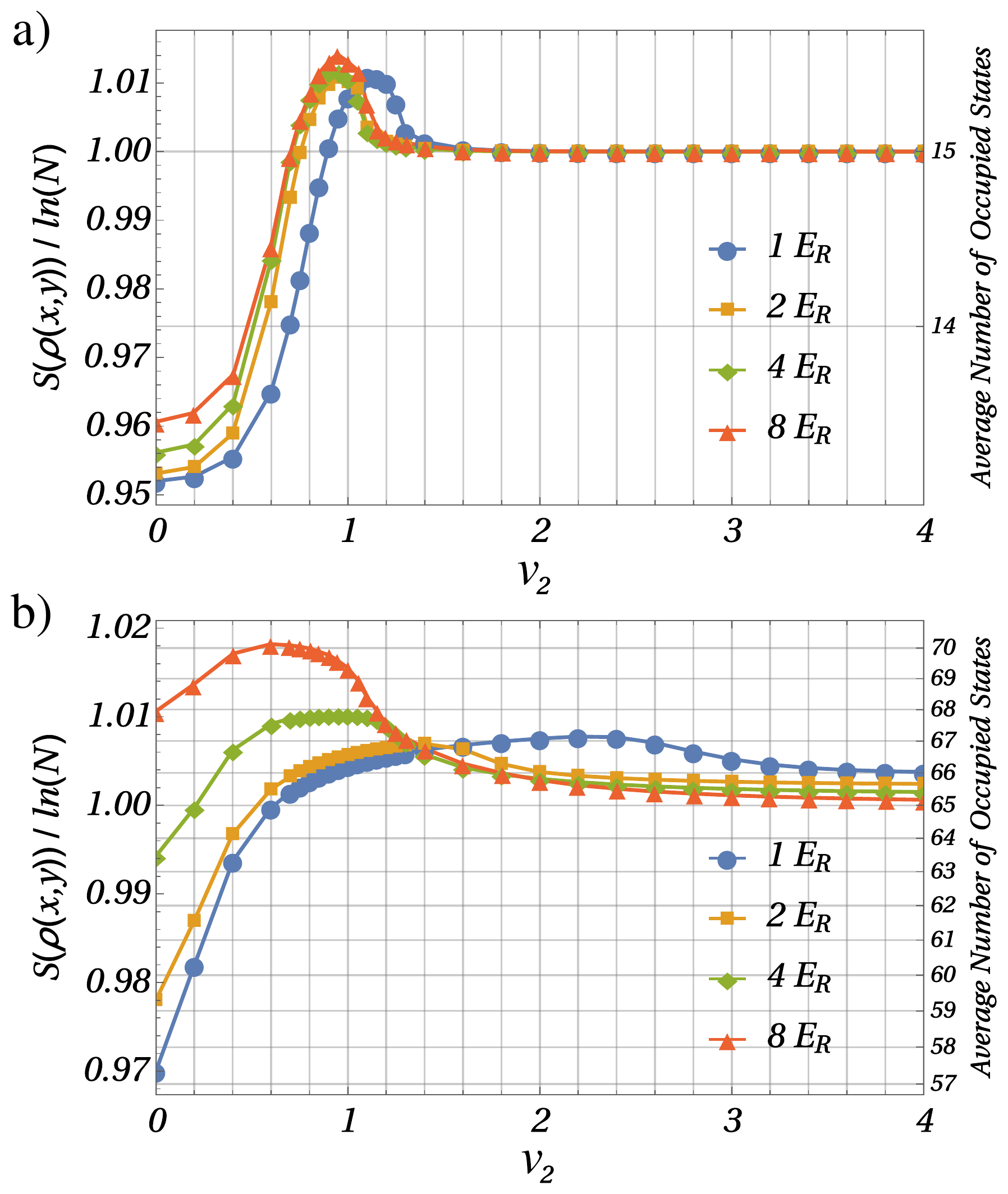}
\caption{(Color online). Entropy of the RSPDM for (top) $N=15$ and
(bottom) $N=65$ bosons in the Tonks-Girardeau regime.}
\label{fig:EntropyRSPDMB}
\end{figure}
However, we expect the entropy to reach such a value in the {\it localized
phase}, where {\it i}) the eigenstates are exponentially localized
in space and, therefore, the strong interaction forbids more than
one boson to occupy an energy level, and {\it ii}) there is a
one-to-one correspondence between energy eigenfuntions and natural
orbitals. As a result, the occupancy of the first $N$ natural
orbitals is $1$, as for fermions, and the entropy is $\ln N$. On the other hand,
in the delocalised phase, the occupation of natural orbitals
changes, with $\lambda_i$ decreasing almost-exponentially with $i$, and
the entropy can take on an arbitrary value $S(\rho_B)>0$. For
weakly interacting bosons at zero temperature, in the superfluid
phase, we have $S(\rho_B) \ll \ln N$ because they all tend
to occupy the same energy level. As the interaction is increased
(but always remaining within the superfluid phase), we expect
bosons to spread in the Hilbert space, resulting in the occupancy
of other natural orbitals, allowing for a more dilute
distribution. This, in turn, leads to $S(\rho_B)>\ln N$. In Fig.
\ref{fig:EntropyRSPDMB} we show the entropy $S(\rho_B)/\ln N$ as a
function of $v_2$ and for different values of the main potential
strength $V_1$ going from the TB to the continuum limits, for the
cases of $N=15$ [Fig. \ref{fig:EntropyRSPDMB} a)], and $N=65$
bosons [Fig. \ref{fig:EntropyRSPDMB} b)]. It can be seen that, as
expected, deep in the localized phase $v_2>1$, the entropy tends
to $\ln N$, showing that bosons tend to occupy one natural orbital
each. It is interesting to compare the behavior of the entropy for
$v_2<1$ for low and large numbers of Bosons. In the first case
(e.g., $N=15$), the entropy is smaller than $\ln 1$ away from the
transition point $v_2\approx 1$, while it exceeds this value
around it. In the second case (e.g. $N=65$), the entropy exceeds
$\ln N$ even in the delocalized phase for $V_1\gg E_r$ (TB limit).
In this limit, indeed, the eigenfunctions are delocalized across
the whole system, but their amplitudes show an increase around the
minima of the main potential (i.e. $V_1$). Therefore two bosons
residing in the same single particle eigenfuction would both be
localized around the minima; as the interaction increases they
naturally tend to occupy other excited states in order to reduce
the average overlap of their wavefunctions. This is why they would
tend to occupy more eigenstates, resulting in a number of occupied
natural orbitals larger than $N$. On the other hand, as the system
is brought in the continuous limit (e.g., for $V_1\approx E_r$),
bosons are allowed to also occupy regions between the minima of
the potential, and therefore the above effect is less important
and the entropy drops below $\ln N$.

Furthermore we can see that the entropy signals the presence of
the ME. To see this we again compare the two cases $N=15$ and
$N=65$ for $V_1=E_r$ (blue circles in Fig.
\ref{fig:EntropyRSPDMB}). In the first case, the entropy rapidly
reaches the asymptotic value $\ln N$, showing the fermion-like
behavior of bosons which occupy one natural orbital each.
Conversely, for $N_p=65$ the asymptotic value is attained for
higher values of $v_2$, showing that some delocalized states are
occupied.

\begin{figure}[h]
\includegraphics[width=7cm]{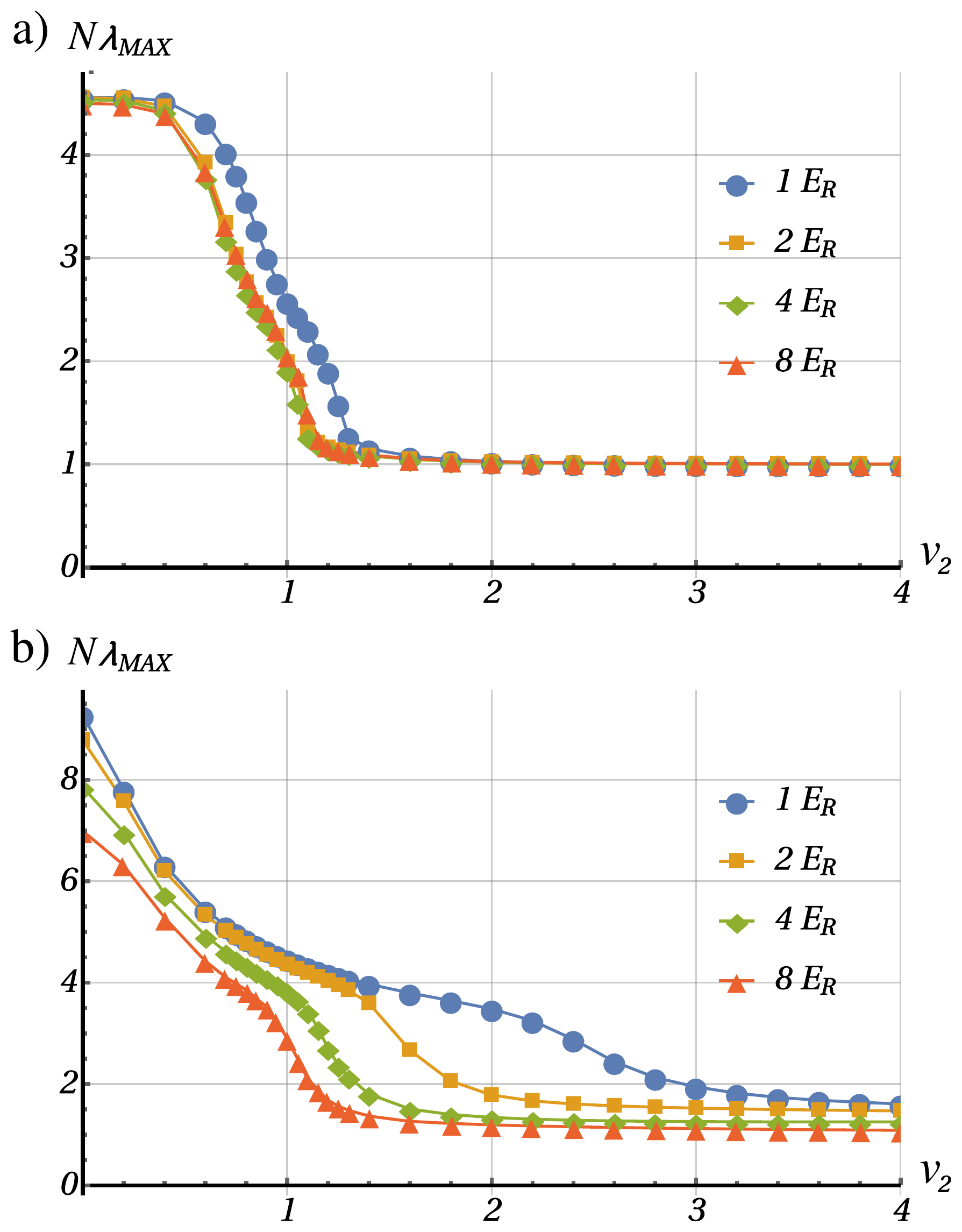}
\caption{(Color online). Fraction of particles in the most largely
occupied natural orbital, given by $N\lambda_{MAX}$ for a) $N_p=15$ and b) $N_p=65$ bosons in
the Tonks-Girardeau regime.} \label{fig:CondensedFractionBosons}
\end{figure}
We can also consider the behavior of the largest eigenvalue
$\lambda_k$ of the RSPDM (Fig. \ref{fig:CondensedFractionBosons}).
As expected, in the localized phase it asymptotically goes to $1$
(as all the first $N$ eigenvalues do), whereas in the delocalized
region it becomes larger than one. Once again the presence of the
ME in the continuum is witnessed by the fact that $\lambda_k$ does
not decay suddenly as a function of $v_2$ (blue circles in Fig.
\ref{fig:DescretePeaksMomDistB} b)).

\section{Conclusions}
We have studied the many-body ground state properties of a system
of non-interacting fermions and strongly interacting bosons in a
one-dimensional bichromatic potential. In the tight-binding regime
we have seen that signatures of the transition are clearly
manifested in the many-body properties of both systems. Similarly,
the presence of a mobility edge in the continuum changes the
many-body properties of the ground state, as shown by comparing
the momentum distribution for different numbers of particles in
the system. If the number of particles is such that only levels
below the mobility edge are filled, then the behavior of the
system is similar to that in the tight-binding regime, as all
occupied states suddenly localize. On the other hand, an increase
in the number of particles results in the mobility edge crossing
the region of the occupied states as the second potential is
varied. This is clearly visible in the momentum distribution of
the system and in the entropy of the reduced single particle
density matrix. Moreover in the case of bosons we have shown that
the interaction plays a key role in the localization properties of
the system.
\section{Acknowledgments}
NLG and FP acknowledge financial support from the EU collaborative project QuProcS
(Grant Agreement 641277).


\end{document}